\documentclass[11pt]{article}
\usepackage{amsmath}
\usepackage{amssymb}
\usepackage{graphicx}
\usepackage{geometry}
\usepackage{bm}
\makeatletter
\renewenvironment{thebibliography}[1]
     {\subsubsection*{References}\footnotesize
      \@mkboth{\MakeUppercase\refname}{\MakeUppercase\refname}%
      \list{\@biblabel{\@arabic\c@enumiv}}%
           {\settowidth\labelwidth{\@biblabel{#1}}%
            \leftmargin\labelwidth
            \advance\leftmargin\labelsep
            \@openbib@code
            \usecounter{enumiv}%
            \let\p@enumiv\@empty
            \renewcommand\theenumiv{\@arabic\c@enumiv}}%
      \sloppy
      \clubpenalty4000
      \@clubpenalty \clubpenalty
      \widowpenalty4000%
      \sfcode`\.\@m}
     {\def\@noitemerr
       {\@latex@warning{Empty `thebibliography' environment}}%
      \endlist}
\makeatother 
\begin{document}
%
\begin{center}
    {\LARGE\bf 
Deterministic and stochastic aspects of the transition to turbulence}
\end{center}
\begin{center}
  Baofang Song$^{1,2}$ and Bj\"orn Hof$^{1,2}$ \\
\end{center}
%
%
\begin{center}
{\small
$^{1}$ IST Austria (Institute of Science and Technology Austria), 3400 Klosterneuburg, Austria\\
$^{2}$ Max Planck Institute for Dynamics and Self-Organization, Bunsenstrasse 10, G\"ottingen, Germany}\\

\end{center}
\vspace{0.2cm}
%
%
\begin{center}
{\bf Abstract}\\
\end{center}
The purpose of this contribution is to summarize and discuss 
recent advances regarding the onset of turbulence in shear flows. 
The absence of a clear cut instability mechanism, the spatio-temporal 
intermittent character and extremely long lived transients 
are some of the major difficulties encountered in these flows and 
have hindered progress towards understanding the transition process. 
We will show for the case of pipe flow that concepts from nonlinear dynamics and statistical 
physics can help to explain the onset of turbulence. In particular 
the turbulent structures ('puffs') observed close to onset are 
spatially localized chaotic transients and their lifetimes increase 
super exponentially with Reynolds number. At the same time 
fluctuations of individual turbulent puffs can (although very rarely) 
lead to the nucleation of new puffs. The competition between these 
two stochastic processes gives rise to a non-equilibrium phase 
transition where turbulence changes from a super-transient to a 
sustained state.\\
\section{Introduction}{\label{sec:intro}}
\indent How turbulence first arises in simple shear flows has remained 
an open question for well over a century. Osborne Reynolds \cite{reynolds1883} was 
the first to observe that this transition depends on a dimensionless 
group, i.e. the Reynolds number, as well as on the amplitude of 
disturbances present in the system. Some of the leading theorists 
at the time (e.g. Lord Kelvin, Lord Rayleigh, Arnold Sommerfeld, 
Werner Heisenberg, Hendrik Antoon Lorentz \cite{Eckert2010}) attempted to probe the stability of pipe and 
related shear flows (i.e. channel and Couette flow) with essentially 
linear methods. After many unsuccessful attempts it has 
become clear (e.g. see \cite{Drazin_Reid1981}) that the occurrence of 
turbulence is unrelated to the stability of the laminar state, as 
Reynolds had already concluded from his experimental observations 
many years earlier. While pipe and Couette flow are believed to 
be stable for all Reynolds numbers, plane Poiseuille (i.e. channel) 
flow becomes unstable at Re of about 5800. However in the latter 
case turbulence is typically already observed at Reynolds numbers 
a little above 1000 and to hold the flow laminar up to the linear
stability threshold actually requires considerable effort in experiments.\\
\indent Transition in the type of flows discussed above is 
qualitatively different from the classical pictures for the transition 
to turbulence which goes back to Landau and to Ruelle and Takens \cite{ruelletakens}. 
In both scenarios turbulence arises following a sequence of 
instabilities of the base flow (which hence becomes linearly unstable).  
While the Ruelle Takens type transition 
has been observed in several closed flows (e.g. \cite{Gollub_Swinney1975}) 
it is noteworthy that even in this case the transition sequence only 
explains the onset of comparably low dimensional chaotic motion 
which dynamically is still far from the full spatio-temporal complexity encountered 
in turbulent flows. In open shear flows such as pipes all experimental 
observations show that the transition in contrast is rather abrupt 
directly from laminar to turbulent. The 
latter type of transition has turned out to be far more difficult to understand. \\
\indent In more recent years a new transition mechanism has been proposed 
based on dynamical systems concepts. Invariant solutions of the 
Navier Stokes equations such as periodic orbits or traveling waves 
are deemed to be ultimately responsible for the existence of the 
turbulent state. These new solutions 
\footnote{The first such solution was discovered for Couette 
flow by Nagata in 1990 \cite{Nagata1990}.}
arise as the Reynolds number 
is increased and importantly are entirely disconnected form the 
laminar flow. The proposition is then that chaotic and ultimately 
turbulent motion will arise following instabilities of these 
disconnected solutions (independent of the laminar state).   Over 
the last two decades or so many traveling waves and periodic 
orbits have been found in direct numerical simulations for 
shear flows (for reviews see \cite{Kerswell2005, Eckhardt2007,Kawahara2012}. 
While coherent structures resembling travelling waves have also been 
observed in turbulent flows in experiments \cite{Hof2004, Hof2005,Lozar2012} 
a main discrepancy remains: close to onset turbulence is always confined 
to structures localized in the streamwise direction and surrounded by laminar 
flow, called ‘puffs’ for the case of pipe flow. Invariant solutions in this
$Re$ range on 
the other hand are usually periodic in this direction. Only very recently 
the first streamwise localized invariant solutions were discovered for 
pipe flow\cite{Avila2013}. In addition it could be shown that chaotic motion indeed 
originates from one such localized periodic orbit. Albeit in this study 
the dynamics were limited to a symmetry subspace it nevertheless 
illustrates how turbulence can arise from a simple invariant solution, 
unrelated to the laminar state, as proposed above. Also this study could 
explain the origin of another property of localized puffs in pipe flow 
which is their transient nature. How (and if) turbulence develops from a 
transient to a sustained state has been subject to much recent debate. 
As we will argue below spatial aspects are crucial in the transition 
mechanism and eventually lead to a non equilibrium phase transition 
giving rise to sustained turbulence. In the following we will discuss 
this transition in more detail and briefly review important recent results.\\
\section{Discussion}
\begin{figure}
\begin{center}
\includegraphics[width=0.8\linewidth]{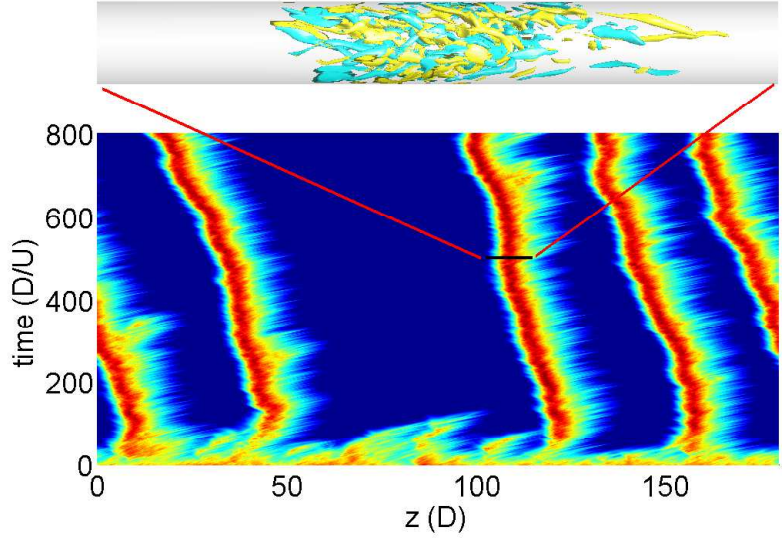}
\caption{\label{fig:puff_spacetime}\small (top) Isosurfaces of the streamwise
vorticity of a puff at $Re=2200$, taken from the segment denoted by a 
black horrizontal bar in the bottom figure. Flow is from left to right. 
(bottom) Space-time diagram for the reduction from $Re$=2800 to 2200 
in a comoving frame with the mean flow. The local turbulence 
intensity $q(z)=\int\int(u_r^2+u_{\theta}^2)rdrd\theta$ 
is plotted versus aixal position in a logrithmic color scale. }
\end{center}
\end{figure}

\begin{figure}
\begin{center}
\includegraphics[width=0.8\linewidth]{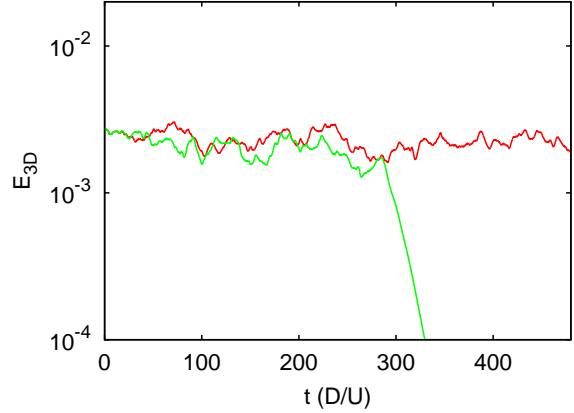}
\caption{\label{fig:Re1850_sensitive}\small Sensitive dependence on initial
conditions of a puff at $Re=1850$. These two lines are the time traces
of the kinetic energy $E_{3D}$ of two runs starting with two very close
initial conditions separated by $\sim 10^{-10}$. While one decays after
about 300 time units, the other persists. Eventually the second one will
also decay (not shown in the figure) due to the transient nature of 
puffs}
\end{center}
\end{figure}

\begin{figure}
\begin{center}
\includegraphics[width=0.8\linewidth]{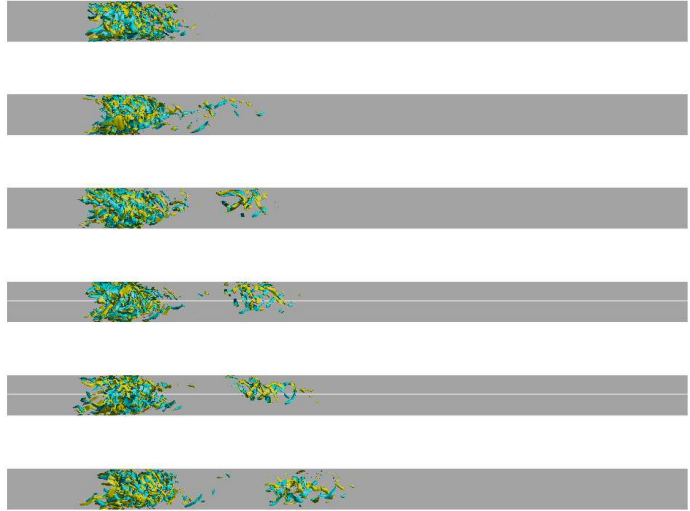}
\caption{\label{fig:splitting}\small Puff splitting process.  At the puff leading edge vortices are shed in the downstream direction.  While normally they decay, occasionally in rare cases they manage to escape beyond the interaction distance of the upstream puff. The first puff and the vortex patch are now separated by a region of laminar fluid and the vortex patch grows to a new puff. The isosurfaces correspond to the axial vorticity component (positive in blue, negative in yellow). }
\end{center}
\end{figure}

\indent In pipe flow turbulent puffs (Fig.~\ref{fig:puff_spacetime}(top)) are typically observed in a 
Reynolds number regime of approximately $1700\lesssim Re\lesssim 2300$. 
They result from perturbations of finite amplitude and in experiments if 
no great care is taken they will often result from distortions the 
flow experiences directly at the inlet. If the inlet is designed 
carefully to avoid such disturbances and if the pipe is sufficiently straight and smooth, flows can be held laminar up 
to much higher $Re$ ( the record in experiments is currently at $Re$=100000). 
For controlled studies of transition it is desirable to start with a 
laminar flow and to study its response to a well controlled disturbance 
which for example can be a jet of fluid injected for a brief period 
through a small hole in the pipe wall. If the amplitude of such a 
perturbation is large enough a turbulent puff is created. While directly 
after the flow has been perturbed the dynamics depend on the nature of the 
disturbance (the puff first has to develop), after 100 to 150 advective time units (measured in D/U, where D is the diameter and U the mean velocity)  the resulting puff is independent of the perturbation that 
triggered it. \\
\indent Curiously in this Reynolds number regime turbulence always 
remains localized. Even if an extended (axially) part of the pipe 
is disturbed the flow will always arrange itself in turbulent 
segments (i.e. puffs) which are about $5$  to $10 D$ long (the turbulent core excluding the leading edge) and interspersed by 
laminar fluid. Vorticity isosurfaces of a turbulent puff at $Re=2200$
are shown in Fig.~\ref{fig:puff_spacetime}(top). 
Fig.~\ref{fig:puff_spacetime}(bottom) shows energy levels in a space time 
plot (time from bottom to top) of a simulation of an initially (t=0) fully 
turbulent flow. Upon reduction of $Re$ to 2200 laminar regions (blue) 
appear and turbulence becomes confined to localized regions, i.e. 
puffs (red vertical stripes in Fig.~\ref{fig:puff_spacetime}(bottom)). 
The direct numerical simulations were carried out with a spectral code 
\cite{Willis2009}. Fourier modes in the axial and azimuthal direction and finite 
difference in the radial are employed and the resolution chosen here is
$48\times(\pm2048)\times(\pm40)$ in radial, axial, and azimuthal directions. 
The domain is $180D$ in the axial direction ($D$ being the diameter) 
with periodic boundary conditions. 
As pointed out in 
\cite{Hof2010} at Reynolds numbers close to 2000 
turbulent puffs extract energy from the adjacent (actually upstream) 
laminar parabolic flow. The plug like turbulent profile in the central 
part of the puff is unable to sustain turbulence (or rather to feed 
turbulence in the downstream direction) consequently the turbulence 
intensity decreases along the leading edge of the puff in the 
downstream direction and the flow eventually relaminarises. Due to 
the action of viscosity the laminar profile now begins to recover 
its parabolic shape so that at sufficient distance a second puff 
can be sustained. If on the other hand the distance between puffs 
is too short the downstream puff will border onto fluid with a flatter 
plug like profile in the upstream direction. Consequently it cannot extract sufficient kinetic 
energy from the flow upstream and decays as shown in 
\cite{Hof2010}. The interaction distance between two puffs is approximately 
$20D$ \cite{Samanta2011}. As a result turbulent puffs have a minimum spacing
of that same distance. While this argument qualitatively explains 
why fully turbulent flow cannot be sustained at these low Re, the 
energetic aspects of this process are not understood in full detail.\\
\indent A key attribute of puffs is their highly chaotic dynamics. This 
gives rise to a loss of memory and limits the prediction of 
the flows future evolution. To illustrate 
the sensitive dependence on initial conditions we simulated two puffs 
with velocity fields that only deviate by $10^{-10}$. As usual for chaotic 
systems this deviation grows exponentially and as can be seen form 
the energy time series shown in Fig.~\ref{fig:Re1850_sensitive}
the signals notably depart 
and become completely unrelated after about 200 
advective time units $D/U$ (the time the puffs takes to travel 200 pipe 
diameters downstream).  The loss of predictabiliy becomes especially 
clear when one puff suddenly decays (green curve in Fig.~\ref{fig:Re1850_sensitive}) 
while the other continues unchanged (i.e. the average quantities remain unchanged).\\
\indent It is a typical feature of puffs that they live for very long 
times and decay suddenly (see Fig.~\ref{fig:Re1850_sensitive}).  
Extensive statistical studies have shown that the survival probability 
is exponentially distributed and the decay is memoryless. 
This behaviour is in line with the escape from chaotic repellers 
observed in lower dimensional systems \cite{Grebogi1983,
Do2004}. Here a chaotic attractor turns into a chaotic saddle 
after an unstable periodic orbit within the attractor and one on the 
basin boundary collide (unstable-unstable pair bifurcation). Above 
the bifurcation point chaotic transients persist for very long times 
before they eventually decay.\\
\indent A similar scenario has recently been observed for pipe flow\cite{Avila2013}.
In this numerical study the dynamics were confined to a symmetry 
subspace (imposing a mirror and a 2-fold rotation symmetry with 
respect to the pipe axis). While this somewhat simplifies the 
dynamics the flow still exhibits turbulent motion for large enough Re. 
By following the laminar turbulent boundary (the so 
called “edge state” \cite{Schneider2007}, which in this 
case is a localised periodic orbit) to lower 
Re the saddle node bifurcation where the periodic orbits originates 
was reached (at $Re$=1430). The upper branch of this saddle node 
however is a stable periodic orbit(see Fig.~\ref{fig:splitting}) and again localised and 
its length is comparable to that of puffs. 
\begin{figure}
\begin{center}
\includegraphics[width=0.8\linewidth]{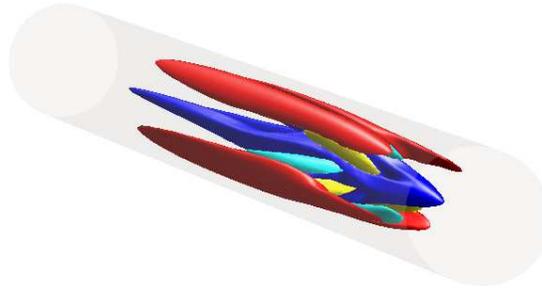}
\caption{\label{fig:orbit}\small Stable periodic orbit in direct numerical simulations of pipe flow with an imposed 2-fold rotation and a reflection symmetry. (Note that this is necessarily also a numerical solution of the full Navier Stokes equation.) This orbit arises in a saddle node bifurcation and is originally stable in the sub-space. At higher Re a bifurcation sequence leads to chaos and transient turbulent puffs. The periodic orbit is shown at $Re=1490$ ( where $Re=UD/\nu$, $\nu$ is the kinematic viscosity, $D$ the pipe diameter and $U$ the mean velocity). Isosurfaces in red/blue show velocity deviations of $+/- 0.1U$ from the laminar parbolic profile. Yellow/ cyan mark isosurfaces of positive/negative streamwise (i.e.axial) vorticity. Upstream and downstream of the periodic orbit the velocity filed quickly approaches the laminar parabolic profile.  }
\end{center}
\end{figure}

For increasing Re the orbit 
first undergoes a secondary Hopf bifurcation followed by the formation 
of a chaotic attractor. This is the point where chaotic motion originates 
in this subspace. The basin of the attractor increases rapidly 
with $Re$ until it reconnects with the unstable periodic orbit on the 
basin boundary. At this moment the chaotic dynamics turn into long 
lived transients (presumably following an unstable-unstable pair bifurcation). 
The memoryless nature of the decay and the loss of predictability is a 
direct consequence of the sensitive dependence on initial conditions 
characteristic for deterministic chaos.\\
\indent  While the classical 
picture of turbulence is that of a chaotic attractor, dating back 
to the landmark paper of Ruelle and Takens, this approach only 
takes the temporal dynamics into account and neglects the spatial 
complexity which is however intrinsic to turbulent flows. 
The importance of spatial aspects and the spatio temporal 
intermittent character of the turbulence transition have been 
emphasized in studies of model system \cite{Kaneko1985, Chate1987}.
Also the chaotic attractor hypothesis has been questioned in 1980's
by Crutchfield and Kaneko \cite{Crutchfield} who propose that chaotic super transients 
are more relevant to turbulence. A first observation 
supporting this view came from direct numerical simulations of pipe 
flows where transients were observed \cite{Brosa1989}. A number of later 
studies were concerned with long lived transients at low Reynolds 
numbers in pipe flow \cite{Faisst2003, Peixinho2006, Hof2006, 
Willis2007, Hof2008, Avila2010, Kuik2010} 
and found that the decay is a memoryless process with a characteristic lifetime $\tau$ which is a function of Re.  There was however 
no consensus if or if not the lifetimes of individual puffs became 
infinite or remained finite. A number of studies proposed that the 
turbulent flow decay rates (inverse characteristic lifetimes) scale as 
$\tau^{-1}\sim(Re_c-Re)$ \cite{ Willis2009,Faisst2003, Peixinho2006} 
or at least as a power law 
$\tau^{-1}\sim(Re_c-Re)^n$ \cite{Willis2009}. 
This view was questioned by Hof \cite{Hof2006} who found that instead 
$\tau^{-1}\sim exp(Re)$ implying that turbulence remains transient. In a 
refined study for much longer observation times than in any previous 
study (spanning almost 8 orders of magnitude in time)  $\tau^{-1}$ was found 
to scale super exponentially with $Re$ \cite{Hof2008}
\footnote{The super exponential scaling has been related to extreme statistics 
theory \cite{Goldenfeld2010}.}.
This scaling was confirmed by 
another experiment \cite{Kuik2010} and in direct numerical simulations 
\cite{Avila2010}. 
\indent It is remarkable that based on their studies 
of spatially coupled maps Crutchfield and Kaneko had not only proposed 
that fluid turbulence could evolve around spatially coupled transients 
but also that the lifetimes under certain conditions (type 2 supertransients) 
are memoryless and increase superexponentially with system size. While it has been 
argued that an increase in Re in turbulent flows is analogous to an increase in system 
size in coupled chaotic maps this correpondence is however not entirely clear. The lifetime 
studies in pipe flow were carried out for single turbulent puffs 
(not many spatially coupled ones) and the puff size hardly changes 
with $Re$. One may argue that the smallest scales of turbulence decrease 
with $Re$ and hence the system size based on this smallest scale increases. 
Nevertheless the Reynolds number range over which the lifetime increase 
is observed is relatively small ($1700<Re<2050$) and hence this size 
effect will be only very moderate. 

\begin{figure}
\begin{center}
\includegraphics[width=0.65\linewidth]{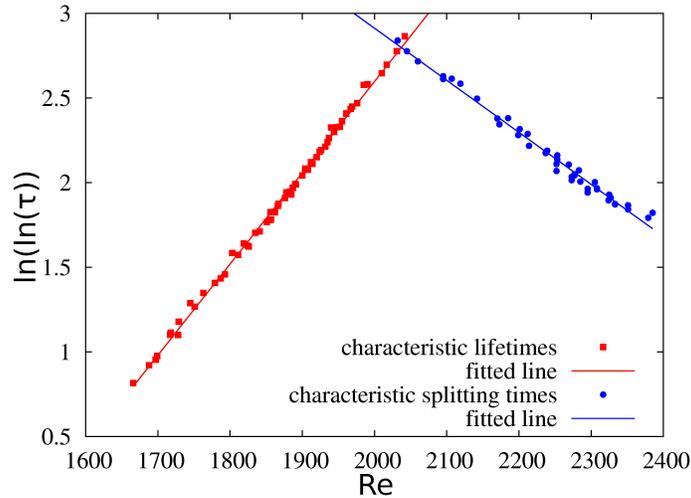}
\caption{\label{fig:decay_split}\small The intersection point between characteristic 
life- and splitting times determines the onset of sustained turbulence in pipe
flow. To the right of the intersection point on average new puffs are
created faster (by puff splitting) than existing puffs decay and hence
(in the thermodynamic limit) spatio-temporally intermittent turbulence
persists. The data is taken from \cite{Hof2008,Avila2011}. 
Note that both the decay and the splitting process are
memoryless and described by characteristic timescales (see original
papers for details}
\end{center}
\end{figure}

\indent A more direct analogy can be drawn 
to another model system of coupled chaotic maps which is motivated by 
bistable excitable media \cite{Barkley2011}. A key difference to the above mentioned map models is that here 
the susceptibility of a “laminar” site to perturbations 
from neighbouring chaotic sites (and hence the minimum perturbation 
amplitude to trigger chaotic dynamics at a laminar site), decreases 
with $R^{-1}$ (where $R$ is a the control parameter analogous to the Reynolds number).  
This model input reflects experimental observations of pipe flow where 
the minimum perturbation amplitude to trigger turbulence was found to 
scale with $Re^{-1}$ \cite{Hof2003, Hof2004, Mullin}.
In the model just like in the experiments localized excited states, 
i.e. puffs, with transient lifetimes are observed and with an increase in 
the control parameter lifetimes of individual puffs scale faster than 
exponential and hence remain transient. \\
\indent Consequently in pipe flow the increase in the temporal 
complexity alone does not lead to sustained turbulence. As 
proposed by Moxey and Barkley \cite{Moxey2010} and later explicitly 
shown by Avila et al. \cite{Avila2011} turbulence becomes sustained 
by a spatial growth process called puff splitting. Puff splitting 
is commonly observed at Reynolds numbers of around 2300
\cite{Wygnanski1975,Nishi2008} and while 
here turbulence is still confined to puffs typically $5$ to $10 D$ in length 
puff sizes fluctuate and can occasionally reach larger values. 
In these instances in a small number  of cases a segment of turbulent 
fluid at the leading edge of the puff can escape further downstream
 beyond the puff-puff interaction distance and a new puff develops 
here (see Fig.~\ref{fig:splitting}). This splitting process leads to an increase in turbulent 
fraction. 
As shown by Avila et al. \cite{Avila2011} 
puff splitting is also intrinsically memoryless and can already be 
found at much lower Re as previously expected. The characteristic time
for such an event to occur decreases super-exponentially with Re. 
The argument for turbulence to become sustained is now straightforward. 
If the characteristic time for turbulent puffs to decay is smaller 
than the time for new puffs to be created (i.e. by splitting) turbulence will 
eventually decay. If on the other hand new puffs are created faster 
than existing ones decay in the thermodynamic limit turbulence becomes 
sustained. The critical Reynolds number where turbulence changes from 
a transient to a sustained state can be estimated by the intersection 
point of the characteristic time scales of the two processes shown in 
Fig.~\ref{fig:decay_split}. \\
\indent This transition is analogous to non equilibrium phase 
transitions and as we will argue below, bears close resemblance to directed percolation (DP) 
and related contact processes. First speculations about a possible 
connection between transition in linearly stable shear flows and DP 
date back to Pomeau in 1986 \cite{Pomeau1986}.
Just like in DP, pipe flow has a unique absorbing state which is 
the laminar flow. Once turbulence has decayed the flow cannot by 
itself return to turbulence unless it is disturbed from the outside. 
The recent studies of puff decay and splitting \cite{Hof2008,Avila2011} 
also suggest that the interaction is only short range, 
which is another requirement for DP  \cite{Hinrichsen2000}. No observations 
indicate that a localized puff would create a second one in a part of 
the pipe not adjacent to it (i.e. more than 25 D or so away). Equally 
it has never been observed that puffs would influence the lifetimes of other 
puffs that are sufficiently far away. These recent studies also infer 
that if such a realtion to DP exists a single unit (i.e. a lattice 
point in DP) must correspond to a turbulent puff and not for example 
to a single vortex. To explore this analogy further hence requires 
much larger system sizes allowing to follow the evolution of many turbulent puffs(/spots). While a number of earlier studies (e.g. \cite{bottin}) looked at such aspects retrospectively the system sizes used were too small. 
In a numerical investigation \cite{Shi2013} of plane Couette we consequently 
chose a narrow but very long domain so that a large 
number of turbulent stripes (analogous structure to puffs in pipes) 
could be accommodated. This study gave more direct evidence that 
the transition is indeed a non-equilibrium continuous phase transition.
Here just like in pipe flow super-exponential lifetime and splitting 
statistics were observed for single stripes, and a critical point for the onset of 
sustained turbulence could be determined in the same manner described above.
The much shorter time scales at the intersection point between 
decay and splitting curves allowed to resolve size distributions 
at this point. The distributions of laminar gaps exhibit scale 
invariance supporting the proposition of a continuous phase transition.
Also here the same transition scenario between transient localized 
chaos and sustained spatio temporal intermittent chaos was found.\\
\indent Analogies to DP have also been explored in recent theoretical studies, 
e.g. \cite{Barkley2011, Sipos2011,Allhoff2012} 
In particular for the coupled map model presented in \cite{Barkley2011}
close agreement was found to the experimental results for pipe flow: the turbulent state becomes sustained when the splitting outweighs the decay of individual puffs. 
In addition the critical exponent for the increase of the turbulent fraction
above onset was found to agree well with the 
universal one for DP in $1+1$ dimension. \\
\indent While at present a final answer to the question if the laminar 
turbulence transition is a non-equilibrium phase transition in accordance 
with DP is outstanding experiments and numerical simulations to clarify 
this question are under way. One of the main challenges here is to 
resolve the extremely long time scales relevant in the vicinity of 
the transition point (note that characteristic splitting and decay 
times in pipes correspond to almost $10^8$ advective time units!). 
In this time puffs trave a distance correpsonding to $10^8$ pipe diameters.  
Equally an accuracy in the Reynolds number of about 0.1\% is 
required setting a further challenge for experiments.  A further 
open issue is the transition from spatially intermittent turbulence 
(i.e. puffs) to expanding space filling turbulent structures which 
takes place somewhere between Reynolds numbers 2300 and 3000.

\end{document}